# Low Voltage Reversible Electrowetting Exploiting Lubricated Polymer Honeycomb Substrates

Edward Bormashenko[1], Roman Pogreb[1], Yelena Bormashenko[1], Roman Grynyov[1], Oleg Gendelman[2]

[1]*Ariel University, Physics Faculty, 40700, P.O.B. 3, Ariel, Israel*

[2]*Faculty of Mechanical Engineering, Technion – Israel Institute of Technology, Haifa 32000, Israel.*

**Abstract**

Low-voltage electrowetting-on-dielectric scheme realized with lubricated honeycomb polymer surfaces is reported. Polycarbonate honeycomb reliefs manufactured with the breath-figures self-assembly were impregnated with silicone and castor oils. The onset of the reversible electrowetting for silicone oil impregnated substrates occurred at 35 V, whereas for castor oil impregnated ones it took place at 80 V. The semi-quantitative analysis of electrowetting of impregnated surfaces is proposed.

An interest in the phenomenon of electrowetting was boosted in the 1980s in the context of various applications of the effect including lab-on-chip systems[1-4] and adaptive optical lenses.[5-7] Numerous applications of electrowetting were summarized in recent reviews.[8-9] The applications of electrowetting face a serious problem: the voltages necessary for manifestations of this effect are relatively high, on the order of magnitude of several hundred volts.[8-9] The papers reporting low-voltage electrowetting are still scanty.[10-11] One of the most popular modern configurations of electrowetting experiments is the so-called electrowetting-on-dielectric scheme (EWOD), in which liquid is placed on an insulating layer on a top of bare electrodes.[12-14] We demonstrate that the EWOD scheme provides low-voltage electrowetting, when lubricated (impregnated) polymer substrates are used as an insulating layer, as depicted in Fig. 1. Aluminum planar electrodes were coated with honeycomb polycarbonate (PC) films, by the fast dip-coating process. As a result, we obtained typical "breath-figures" self-assembly patterns, depicted in Fig. 2. Honeycomb PC coating was obtained according to the protocol described in detail in Refs. 15-16. The average radius of pores was about 1.5 μm. The average depth of pores as established by AFM was about 1 μm. PC porous coatings were impregnated by two kinds of oils: castor oil (supplied by Vitamed Pharmaceutical Industries, LTD) and silicone oil for MP&BP apparatus, for brevity called hereafter "silicone oil" (supplied by Sigma-Aldrich). For the purpose of uniform spreading of oils, the lubricated substrates were heated to 60°C, and afterwards cooled to ambient temperature. The thickness of both oil layers was established by weighting as 20±2 μm. The physical properties of the oils relevant to our study are summarized in Table 1. Experiments were carried out with 8 μl bi-distilled water droplets. Droplets were visualized with a Ramé-Hart Advanced Goniometer, Model 500-F1.

The wetting properties of polymer substrates impregnated by oils were studied systematically by Aizenberg et al.[17-19] Theoretical approaches to the wetting of impregnated substrates were developed by Cohen, Nosonovsky et al.[20-23] Impregnated polymer substrates exhibit extremely low contact angle hysteresis.[17] This fact makes them suitable candidates for the low-voltage electrowetting, due to the weak pinning of the triple line.[23-26] It was shown recently that manufacturing of non-pinning, low contact angle hysteresis surfaces is an appropriate way to develop low voltage driven electrowetting devices.[2, 5, 27-28]

Indeed, PC oil-impregnated honeycomb surfaces demonstrated low voltage DC electrowetting, illustrated by Fig. 3 (representing the relative maximal displacement of the triple line $\Delta D/D_0$, as a function of the applied voltage $U$). Fig. 4 depicts electrowetting of silicone oil impregnated substrates. "Wetting ridges" formed in the vicinity of the triple line, discussed in Ref. 21, are distinctly seen. The ridges were observed for both castor and silicon oils used as lubricants. The complicated shape of the water/vapor interface, exhibiting a flex point, is noteworthy, making accurate measurement and interpretation of the contact angle quite challenging.[21] That is why, we preferred to quantify the electrowetting of impregnated surfaces in terms of the relative maximal displacement of the triple line $\Delta D/D_0$ (see Fig. 3).

It is seen from the data, displayed in Fig. 3 that silicone oil impregnated substrates are more suitable for low-voltage electrowetting than castor oil impregnated ones. The onset of the process for silicone oil impregnated substrates occurred at 35 V, whereas for castor oil impregnated ones it took place at 80 V. Moreover, the sensitivity of the electrowetting scheme to applied voltage is much larger for silicone oil impregnated substrates, as it is clearly seen from Fig. 3. It is reasonable to quantify the sensitivity of the electrowetting scheme by the parameter $\xi = \Delta D/UD_0$. It is recognized that for the silicone oil based EWOD scheme $\xi_{\text{silicone}} \cong 1.3 \cdot 10^{-3} \text{V}^{-1}$; whereas for the castor oil based EWOD, we established $\xi_{\text{castor}} \cong 3.2 \cdot 10^{-4} \text{V}^{-1}$.

This result deserves more extended discussion. In a classical electrocapillarity set-up, the phenomenon of electrowetting is related to formation of the Helmholtz double-layer at the interface between metal and electrolyte.[9, 10, 29] Charges at the interface form a parallel plate capacitor in which the gap thickness is on the order of a Debye-Huckel length.[29] Within the modern EWOD scheme an electrolyte contacts the dielectric layer coating the metal, which is the impregnated PC honeycomb film in our experiments; thus the charges separation is micrometrically scaled.[12-14] The well-known Lippmann Equation governing the electrowetting predicts for the change in the contact angle (for both classical and EWOD schemes):

$$\cos\theta^* = \cos\theta_Y + \frac{\tilde{C}U^2}{2\gamma}, \qquad (1)$$

where $\theta^*$ is the apparent contact angle of electrowetting, $\theta_Y$ is the equilibrium contact angle, $\tilde{C}$ is the specific capacity of the unit area confined by a double layer, and $\gamma$ is the surface tension of water.[8, 9, 25] The traditional Lippmann Equation supplied by Exp. 1, is not straightforward applicable for the analysis of electrowetting of lubricated surfaces, due to the complicated balance of interfacial tensions, illustrated with Fig. 1. However, it is expected from Eq. 1, that a castor oil impregnated PC surface is a better candidate for electrowetting, since its dielectric constant is rather high: $\varepsilon = 4.7$ (see Table 1). The experimental result, however, is opposite. Several factors may be responsible for this. First of all, the surface tensions of castor and silicon oils and interfacial tensions of oil/water systems differ markedly (see Table 1). This difference may lead to very different regimes of wetting, as discussed in detail in Ref. 21. However, we established that the sliding angles for 8 μl droplets deposited on both castor and silicone oils-impregnated honeycomb surfaces were the same, namely 5°. Hence, it is reasonable to suggest that both castor and silicone oils coat the water droplet, as it is shown in Ref. 21. This may be shown also by the analysis of the parameter $S$ governing the spreading: $S = \gamma - (\gamma_{oil} + \gamma_{oil/water})$, where $\gamma, \gamma_{oil}, \gamma_{oil/water}$ are interfacial tensions at water/vapor, oil/vapor and oil/water interfaces respectively. Interfacial tensions supplied in Table 1 are taken from the literature data.[30,31]

Assuming $\gamma = 71 \frac{mJ}{m^2}$, we obtain $S > 0$ for both castor and silicone oils, which in this case are expected to coat the water droplet.[32]

The second factor, which may explain the high sensitivity of the silicone oil based EWOD scheme to applied voltage, is related to the relatively low viscosity of silicone oil, which is much lower than that of the castor oil (see Table 1).

Consider the semi-quantitative analysis of electrowetting of impregnated surfaces. Figure 5 depicts the average velocity $v$ of the triple line, established experimentally as a function of the applied voltage $U$. The averaged velocity was defined as $v = \Delta D / \tau$, where $\tau$ is the total time of displacement of the triple line. We observe that the characteristic velocity of the triple line is in the range of $0.1 \div 0.6$ mm/s. Hence, the Reynolds number $Re$ can be estimated as: $Re = \rho v D / \eta \approx 2 \cdot 10^{-2} \div 2 \cdot 10^{-4}$. Thus, the inertia forces are negligible. The physical mechanism of the movement of droplets on the lubricated surfaces was treated in detail in Ref. 21. Generally, viscous dissipation under displacement of a droplet occurs in the bulk of the droplet, at the wetting ridge, and within the lubricating layer.[21] In the case of electrowetting, the velocity of the center mass of the droplet is negligible; hence, the viscous dissipations in the droplet bulk and wetting ridge are also negligible.[21] Consequently the de-pinned droplet is stopped by the viscous force developed by the lubricant. Thus, the balance of interfacial and viscous forces acting within the oil layer supplies the following estimation:

$$\Delta \gamma(U) \pi D \cong \eta \frac{\pi D^2}{4} \frac{v}{h}, \qquad (2)$$

where $\Delta \gamma(U) = |\gamma_{oil/water}(U)(1 + \cos\theta^*) - \gamma_{oil}(1 - \cos\theta^*)|$ is the non-equilibrium, specific, interfacial force, depending on the voltage, and $dv/dh \cong v/h$ is assumed

for the purposes of the rough estimation ($h$ is the thickness of the oil layer). Expression (2) yields for the estimation of the triple line velocity:

$$v \cong \frac{4h\Delta\gamma(U)}{\eta D}. \qquad (3)$$

The reasonable estimation of $\Delta\gamma$ is: $\Delta\gamma \cong 5\frac{mJ}{m^2}$; substituting for the castor oil $h \cong 20\,\mu m$, $D \cong 3\,mm$; $\eta \cong 1\,Pa \times s$ yields $v \cong 0.13\,mm/s$, in a satisfactory agreement with the experimental findings. At the same time, Exp. (3) overestimates the velocity of the triple line for the silicone oil/water electrowetting experiment. Perhaps this is due to the overestimation of $\Delta\gamma$ for the silicone oil/water pair. It should be stressed that up to a certain threshold value of voltage (which is 80 V for silicone oil and 120 V for castor oil) the electrowetting of impregnated polymer honeycomb substrates is reversible. We conclude that impregnated porous polymer substrates demonstrate a potential for low voltage electrowetting.

**Acknowledgements**

The Authors are grateful to Professor Gene Whyman for extremely fruitful discussions. The Authors are thankful to Mrs. Albina Musin for her inestimable help in preparing this manuscript.

Table 1. Physical properties of oils used in the investigation.

| Oil | Viscosity, $\eta$, Pa×s | Surface tension, $\gamma_{oil}$, mJ/m$^2$ | Interface tension, $\gamma_{oil/water}$ mJ/m$^2$ | Density, $\rho$, 10$^3$ kg/m$^3$ | Dielectric constant, $\varepsilon$ |
|---|---|---|---|---|---|
| Castor oil | 0.985 | 40.4 | 23-24 | 0.961 | 4.7 |
| Silicone oil | 0.05 | 20 | 41-43 | 0.963 | 2.5 |

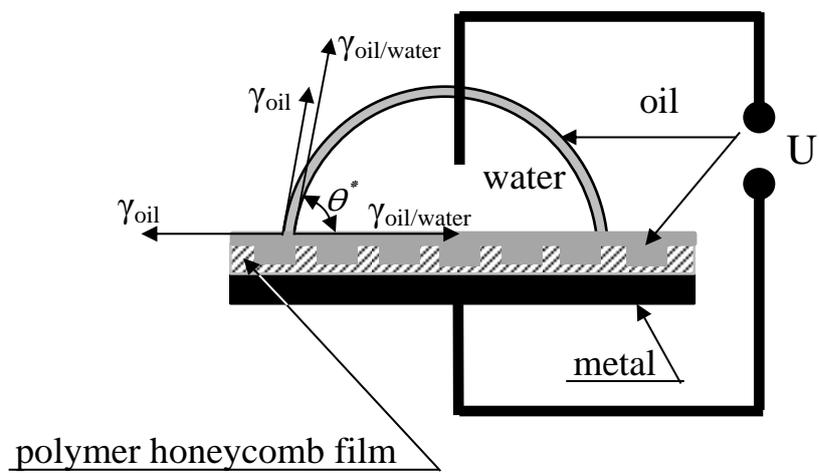

FIG. 1. EWOD scheme exploiting lubricated honeycomb polymer layer as an insulating layer. The balance of interfacial forces is shown.

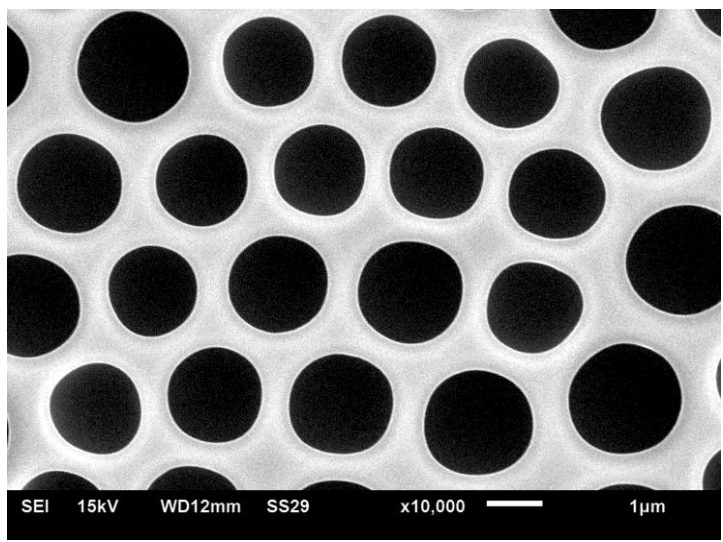

FIG. 2. Polycarbonate honeycomb coating of Al electrodes obtained with "breath-figures", carried out in a humid atmosphere. Scale bar is 1 μm.

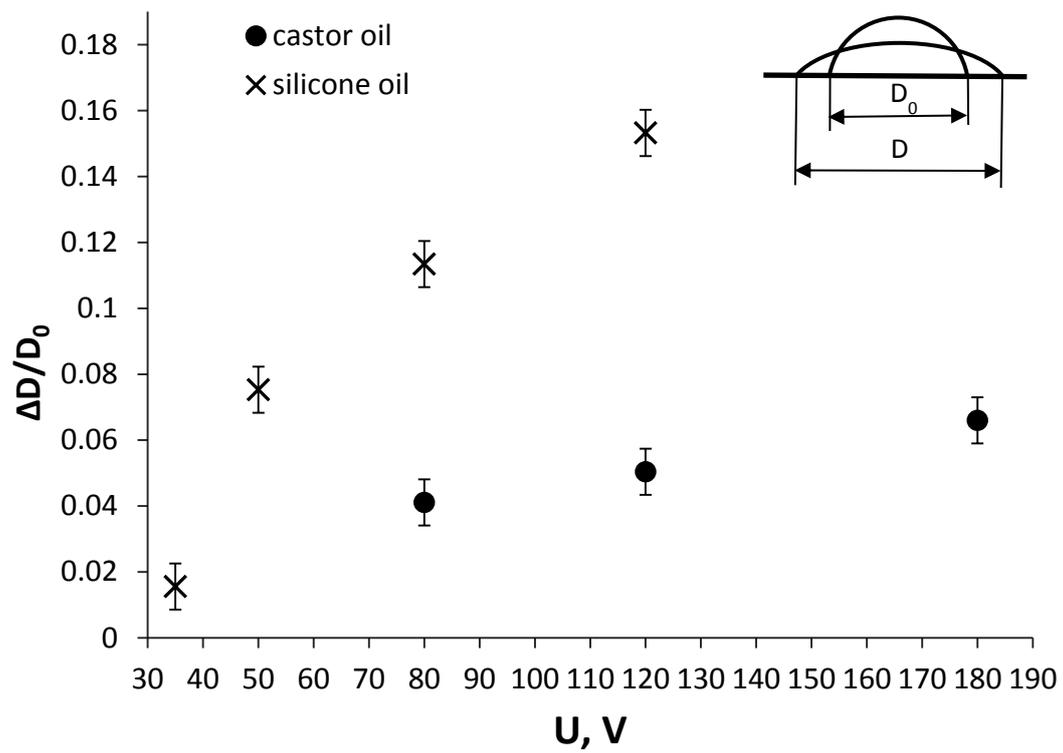

FIG. 3. Relative maximal displacement of the triple line $\Delta D / D_0$ vs. applied DC voltage $U$ for different oils used in the investigation.

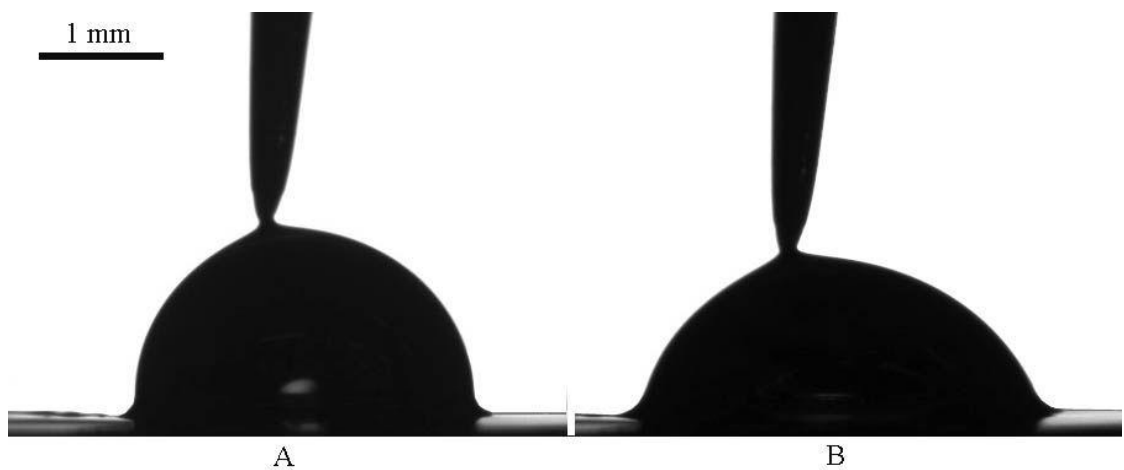

FIG. 4. Electrowetting of silicone oil lubricated PC substrates. Volume of water droplet is 8 μl. (a) $U$=0V; (b) $U$=120 V.

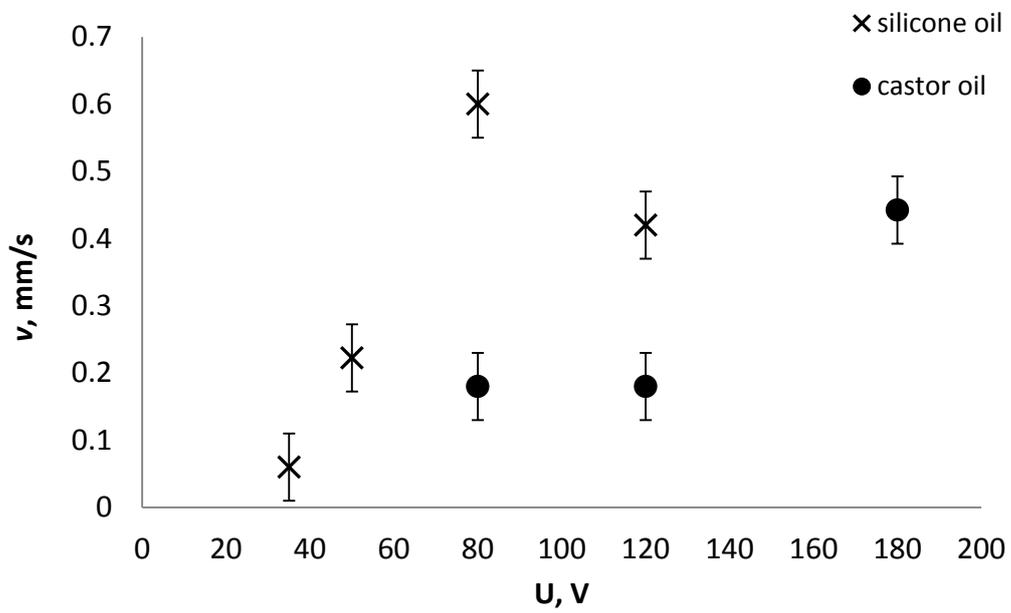

FIG. 5. Average velocity of the triple line *v* vs. applied voltage *U* for the honeycomb substrates, impregnated with silicone and castor oils.